\documentclass[a4paper,fleqn]{cas-sc}

\usepackage{mathtools}
\usepackage{threeparttable}
\usepackage{graphicx}
\usepackage[authoryear,longnamesfirst]{natbib}

\def\tsc#1{\csdef{#1}{\textsc{\lowercase{#1}}\xspace}}
\tsc{WGM}
\tsc{QE}
\tsc{EP}
\tsc{PMS}
\tsc{BEC}
\tsc{DE}

\begin{document}
\let\WriteBookmarks\relax
\def\floatpagepagefraction{1}
\def\textpagefraction{.001}
\shorttitle{GARDRec: Decision-Level Graph Grounding for LLM Recommendation}
\shortauthors{Yong Wang et~al.}

\title [mode = title]{GARDRec: Decision-Level Graph Grounding for Large Language Model Recommendation}

\author[2]{Yong Wang}
\fnmark[1]
\ead{wangyonghao114@gmail.com}

\author[1,2,3]{Hongliang Sun}[orcid=0000-0002-5333-1378]
\fnmark[1]
\ead{sunhl@hit.edu.cn}

\author[2]{Jinlan Liu}[orcid=0000-0002-5333-1378]
\ead{25S130324@stu.hit.edu.cn}

\author[2]{Hua Zhang}
\cormark[1]
\ead{zhua@hit.edu.cn}

\author[2]{Dianbo Sui}
\ead{suidianbo@hit.edu.cn}

\author[2,3]{Dianhui Chu}[orcid=0000-0003-2973-7252]
\ead{chudh@hit.edu.cn}

\author[2,3]{Zhiying Tu}[orcid=0000-0001-8800-4513]
\cormark[1]
\ead{tzy_hit@hit.edu.cn}

\fntext[fn1]{Yong Wang and Hongliang Sun contributed equally to this work.}

\cortext[cor1]{Hua Zhang and Zhiying Tu are the Corresponding authors.}


\affiliation[1]{organization={Harbin Institute of Technology Qingdao Research Institute},
                city={Qingdao},
                postcode={266109},
                country={China}}

\affiliation[2]{organization={Harbin Institute of Technology (Weihai)},
                city={Weihai},
                postcode={264209},
                country={China}}

\affiliation[3]{organization={Shandong Provincial Key Laboratory of Digital Service Computing Technology and Systems},
                city={Weihai},
                postcode={264200},
                country={China}}

\begin{abstract}
Large language models (LLMs) offer new opportunities for recommendation by interpreting item descriptions, user instructions, and external knowledge through natural-language prompts. However, existing graph-augmented LLM recommenders often use knowledge graphs mainly as prompt-level evidence, leaving ranking decisions weakly constrained by structured user--item relations. This is problematic for next-item recommendation, where the model must compare candidates under the same user context while preserving temporal preference, collaborative signals, and attribute matches. To address this issue, we propose \emph{GARDRec}, a Graph-grounded Adaptive Reasoning and Decision-aware Recommendation framework for LLM-based next-item ranking. GARDRec constructs semantic-structural item representations from textual node features and graph propagation, derives personalized graph contexts from temporally weighted histories and first-order neighborhoods, and aligns graph-derived representations with a frozen LLM through continuous multimodal prompts. Explicit interaction and matching features are injected through late-stage decision branches, while inter-candidate attention and restricted generative likelihood support final ranking. Experiments on three public benchmarks with multiple LLM backbones show that GARDRec generally improves candidate-ranking performance over representative baselines. Ablation and diagnostic analyses verify the contributions of graph projection, neighborhood retrieval, explicit decision features, ranking loss, and generative calibration.
\end{abstract}


\maketitle

\section{Introduction}

Recommender systems have become a central interface for navigating large-scale digital content, products, and services. Classical collaborative filtering, matrix factorization, and neural collaborative filtering model user--item interaction patterns and make personalized recommendation scalable for implicit feedback scenarios \citep{resnickGroupLensOpenArchitecture1994,korenMatrixFactorizationTechniques2009,heNeuralCollaborativeFiltering2017}. Sequential recommendation further shifts the task from estimating static preference to predicting the next item from recent user behavior, where temporal order, short-term intent, and candidate-level comparison are decisive \citep{kangSelfAttentiveSequentialRecommendation2018,sunBERT4RecSequentialRecommendation2019}. However, interaction-only models remain limited when user histories are sparse, items contain rich textual attributes, or recommendation decisions require semantic evidence beyond observed clicks.

Graphs provide a natural abstraction for recommendation because users, items, attributes, and interactions are inherently relational. Graph neural recommenders propagate collaborative signals over user--item graphs and item knowledge graphs, improving the modeling of high-order connectivity, long-tail items, and side information \citep{yingGraphConvolutionalNeural2018,wangNeuralGraphCollaborative2019,heLightGCNSimplifyingPowering2020,wuSelfsupervisedGraphLearning2021}. Knowledge-graph-based recommendation further incorporates entities and relations such as actors, genres, authors, and categories, allowing user preferences to be expanded through structured item attributes and paths \citep{wangRippleNetPropagatingUser2018,wangKGATKnowledgeGraph2019}. However, conventional graph recommenders usually compress relational evidence into task-specific embeddings, making it difficult for recommendation models to fully exploit relational semantics for user preference reasoning.

LLMs introduce a different opportunity for recommendation: they can read item descriptions, interpret natural-language instructions, and act as rankers or conversational recommenders under flexible prompts \citep{gaoChatRECInteractiveExplainable2023,baoTALLRecEffectiveEfficient2023,zhangRecommendationInstructionFollowing2023,houLargeLanguageModels2024,luoRecRankerInstructionTuning2024}. Yet LLM-based recommendation also inherits the weaknesses of language-model inference. Autoregressive models are sensitive to prompt order and context placement, may overlook structured dependencies, and can produce recommendations that are semantically plausible but weakly grounded in user--item relations \citep{liuLostMiddleHow2023,jiSurveyHallucinationNatural2023}. Retrieval-augmented generation (RAG) mitigates part of this issue by injecting external evidence into the input context \citep{lewisRetrievalAugmentedGenerationKnowledgeIntensive2021,zhaoRetrievalAugmentedGenerationAIGenerated2024}, and recent graph-augmented LLM methods retrieve triples, paths, or subgraphs to provide structured knowledge for downstream reasoning \citep{baekKnowledgeAugmentedLanguageModel2023a,perozziLetYourGraph2024a,heGRetrieverRetrievalAugmentedGeneration2024a,panUnifyingLargeLanguage2024}. However, how such graph evidence should influence the ranking decision in recommendation remains less clear.

\begin{figure}
	\centering
	\includegraphics[width=0.8\textwidth]{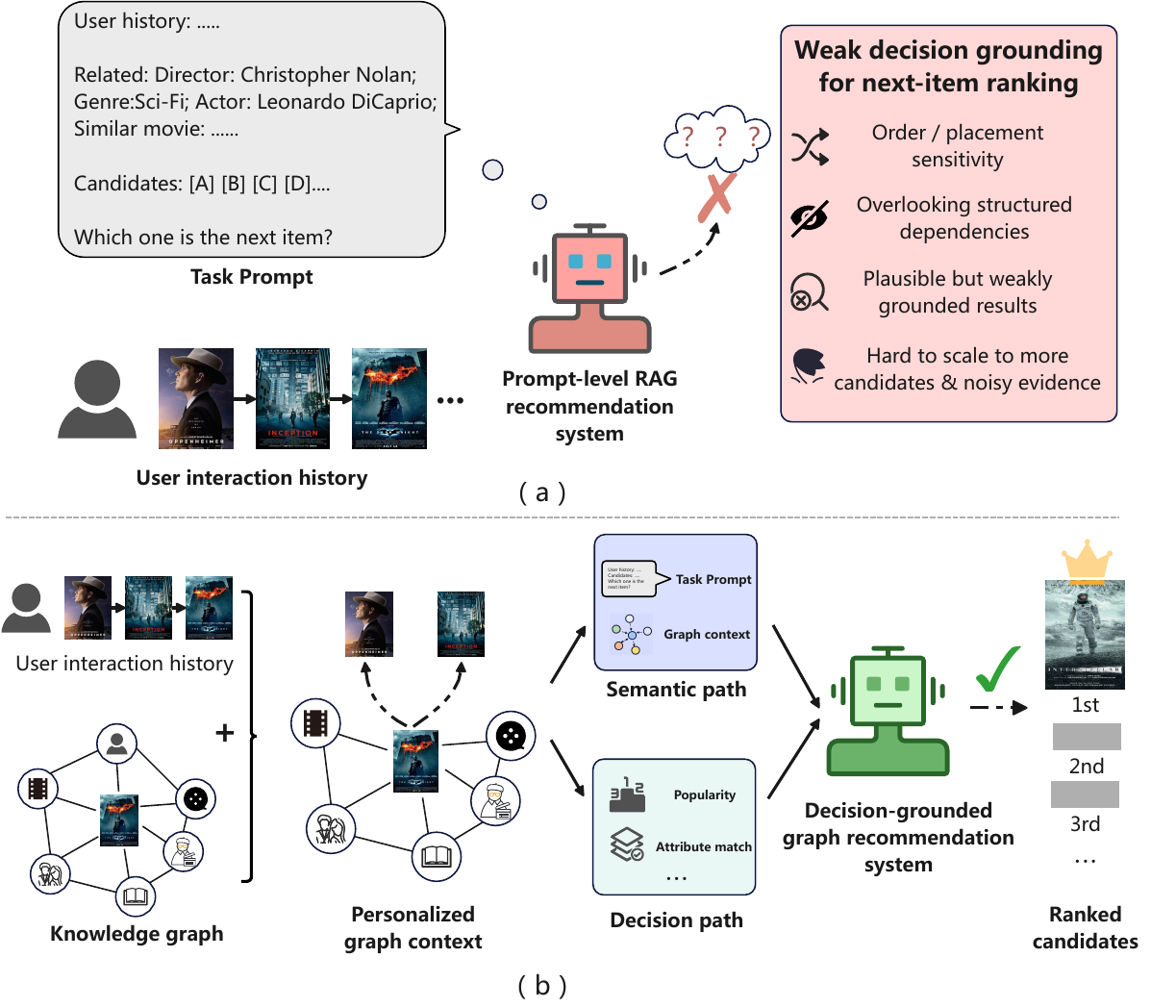}
	\caption{(a) Prompt-level graph augmentation exposes related facts to an LLM, but the final next-item decision can remain sensitive to order, context placement, and weakly modeled structured dependencies. (b) A decision-grounded graph recommender constructs personalized graph context from user history and knowledge-graph neighborhoods, allowing relational evidence to guide candidate comparison before final ranking.}
	\label{fig:introduction_motivation}
\end{figure}

Although retrieval-augmented recommendation has shown promise, treating graphs mainly as prompt-level evidence is insufficient for next-item ranking. As illustrated in Figure~\ref{fig:introduction_motivation}, a prompt-only graph-augmented pipeline can provide useful entities, attributes, and related items, but it still leaves the LLM to implicitly resolve candidate competition and preserve fine-grained relational constraints. This distinction matters because next-item ranking is not merely a knowledge-reading task; it is a constrained comparison task. In multi-candidate recommendation, the model must compare candidates under the same user context rather than score them in isolation. It also needs to preserve temporal asymmetry and fine-grained collaborative signals, such as transition, co-occurrence, popularity, and hard attribute matches. Serializing graph evidence into text or injecting graph tokens into a frozen LLM does not guarantee that these numerical and relational constraints will be preserved during long-context attention, especially when the candidate set grows or retrieved evidence contains noise. Closest KG-RAG methods for LLM-based recommendation, such as K-RagRec, improve knowledge retrieval by selecting high-quality graph information for prompt augmentation \citep{wangKnowledgeGraphRetrievalAugmented2025}; however, the broader question remains underexplored: how can graph knowledge constrain the recommendation decision itself rather than merely enrich the LLM input?

We address this question by proposing \emph{GARDRec}, a \emph{Graph-grounded Adaptive Reasoning and Decision-aware Recommendation} framework for LLM-based next-item ranking. The key insight is that knowledge graphs should serve as a decision-level grounding layer, not only as an external retrieval source. GARDRec first builds semantic-structural item representations by combining pretrained textual node features with graph propagation. It then derives personalized graph contexts from temporally weighted user histories and first-order graph neighborhoods, and aligns these graph-derived contexts with the hidden space of a frozen LLM through continuous prompt representations. Instead of relying solely on generative likelihoods, GARDRec explicitly models inter-candidate competition and uses decoupled scoring branches to preserve hard matching features and collaborative interaction signals outside the text prompt. This design allows graph evidence to participate in representation, context expansion, candidate comparison, and final ranking.

The contributions of this work are summarized as follows:
\begin{itemize}
    \item We formulate graph-grounded LLM recommendation as a decision-level ranking problem, where knowledge graphs serve as structured grounding layers for user intent, candidate relations, and collaborative constraints rather than as prompt-only evidence. 
    \item We propose GARDRec, a graph-grounded LLM recommendation framework that jointly integrates semantic-structural item representation, personalized graph-context construction, multimodal LLM alignment, and decision-aware candidate scoring. 
    \item We conduct extensive experiments on three public recommendation benchmarks with multiple LLM backbones. The results demonstrate that GARDRec achieves stronger performance in most settings, and further analyses verify the effectiveness of its key components.
\end{itemize}

\section{Related Work}

\subsection{Sequential and Graph-based Recommendation}

Recommendation has long been studied through collaborative signals, content signals, and their hybridization. Collaborative filtering, matrix factorization, and neural collaborative filtering model user preference from historical user--item interactions and improve the expressiveness of latent interaction modeling \citep{resnickGroupLensOpenArchitecture1994,korenMatrixFactorizationTechniques2009,heNeuralCollaborativeFiltering2017}. Content-based recommendation complements interaction modeling by exploiting item attributes and textual descriptions \citep{lopsContentbasedRecommenderSystems2011}. For next-item recommendation, sequential models further incorporate the temporal order of user behaviors, with self-attention-based architectures such as SASRec and BERT4Rec showing that recent interactions and contextual dependencies are crucial for candidate ranking \citep{kangSelfAttentiveSequentialRecommendation2018,sunBERT4RecSequentialRecommendation2019}. These methods provide strong foundations for preference modeling, but they generally rely on compact interaction representations and provide limited access to explicit relational evidence for the final decision.

Graph-based recommendation makes relational evidence more explicit by representing users, items, attributes, and interactions as structured graphs. Graph convolutional recommenders propagate signals over user--item graphs to capture high-order collaborative patterns, and large-scale graph models such as PinSage demonstrate the practical value of graph representation learning for web-scale recommendation \citep{yingGraphConvolutionalNeural2018,wangNeuralGraphCollaborative2019,heLightGCNSimplifyingPowering2020}. Self-supervised graph learning further improves robustness by constructing graph-augmented views for recommendation \citep{wuSelfsupervisedGraphLearning2021}. Knowledge-graph recommenders extend this line by connecting items with semantic entities and relations, enabling user interests to propagate over item attributes and high-order paths \citep{wangRippleNetPropagatingUser2018,wangKGATKnowledgeGraph2019}. Our work follows the graph-based view that relational structure is essential for recommendation, but differs from conventional graph recommenders by using the graph as a grounding layer for an LLM-based ranking process rather than only as a task-specific representation module.

\subsection{Large Language Models for Recommendation}

LLMs have introduced a new paradigm in which recommendation can be formulated as instruction following, conversational interaction, or candidate ranking. Early LLM-based recommenders convert user histories and item descriptions into prompts, allowing the language model to use its semantic knowledge and natural-language understanding for recommendation \citep{gaoChatRECInteractiveExplainable2023,zhangRecommendationInstructionFollowing2023,liuChatGPTGoodRecommender2023}. Efficient tuning methods such as TALLRec align LLMs with recommendation data, while instruction-tuned rankers and auto-reranking frameworks adapt LLMs to top-$K$ recommendation and reranking settings \citep{baoTALLRecEffectiveEfficient2023,houLargeLanguageModels2024,luoRecRankerInstructionTuning2024,gaoLLM4RerankLLMbasedAutoReranking2025}. Other studies incorporate collaborative embeddings or graph-derived signals into LLMs to compensate for the weak collaborative grounding of purely textual prompting \citep{zhang2023collm,weiLLMRecLargeLanguage2024,jiangRecLMRecommendationInstruction2025}.

Despite their flexibility, LLM-based recommenders face a structural mismatch between language modeling and ranking. Autoregressive LLMs are sensitive to context order and may underuse information located in the middle of long prompts \citep{liuLostMiddleHow2023}. Their outputs can also be weakly grounded when the model relies on parametric knowledge or surface-level item semantics rather than explicit user--item relations, a concern aligned with broader findings on hallucination in natural language generation \citep{jiSurveyHallucinationNatural2023}. These limitations are particularly important for multi-candidate next-item ranking, where the model must preserve temporal preference, compare candidates under a shared user context, and incorporate precise collaborative signals such as transition, co-occurrence, and popularity. This motivates our design of separating LLM-based semantic reasoning from decision-level graph calibration.

\subsection{Retrieval-Augmented and Graph-Augmented LLMs}

Retrieval-augmented generation provides a general mechanism for grounding LLM outputs in external evidence. RAG models combine parametric generation with retrieved non-parametric memory for knowledge-intensive tasks \citep{lewisRetrievalAugmentedGenerationKnowledgeIntensive2021}, and later studies summarize retrieval augmentation as a broad strategy for improving factuality, freshness, and robustness in generative AI \citep{zhaoRetrievalAugmentedGenerationAIGenerated2024}. Graph-augmented LLMs further exploit structured external knowledge by verbalizing triples, encoding graph structures, or retrieving graph substructures for downstream reasoning \citep{baekKnowledgeAugmentedLanguageModel2023a,wuRetrieveRewriteAnswerKGtoTextEnhanced2023a,perozziLetYourGraph2024a,heGRetrieverRetrievalAugmentedGeneration2024a}. The broader roadmap of LLM--knowledge graph integration also emphasizes that graphs can provide explicit structure and factual grounding for language models \citep{panUnifyingLargeLanguage2024}.

Recent studies adapt retrieval augmentation to recommendation by retrieving different forms of external evidence before LLM inference. Some methods retrieve long-term user behaviors or collaborative evidence to improve sequential and long-tail recommendation \citep{lin2024rella,wu2024coral}. Others focus on item-level evidence or graph-based retrieval for explainable recommendation \citep{kim2025itemrag,li2025grefer}. KG-oriented methods further retrieve or construct knowledge-graph contexts to augment LLM-based recommendation, including K-RagRec, KERAG\_R, and LlamaRec-LKG-RAG \citep{wangKnowledgeGraphRetrievalAugmented2025,meng2025kerag,azizi2025llamarec}. These studies demonstrate the importance of external evidence, but they primarily focus on what information should be retrieved and injected into the LLM context. In contrast, our work focuses on how graph evidence should shape the ranking decision after retrieval.

\subsection{Positioning of This Work}

The closest line of research to this paper is KG-enhanced RAG for LLM-based recommendation. K-RagRec retrieves high-quality structural information from knowledge graphs to augment recommendation generation, and related GraphRAG-style recommenders improve the selection or construction of graph context \citep{wangKnowledgeGraphRetrievalAugmented2025,meng2025kerag,azizi2025llamarec}. GARDRec shares the motivation that LLM recommendation should be grounded in structured knowledge, but it departs from prompt-level augmentation in three aspects. First, it constructs semantic-structural graph representations and personalized graph contexts that can be aligned with the LLM hidden space. Second, it explicitly models inter-candidate competition, which is central to next-item ranking but often remains implicit in generation-style recommendation. Third, it preserves hard matching and collaborative interaction features through decoupled scoring branches, reducing the dilution of precise numerical signals inside long textual prompts. This positioning makes GARDRec complementary to retrieval-focused RAG methods: retrieval decides what evidence is available, whereas decision-level graph grounding decides how that evidence shapes the final ranking.

\section{Methodology}

This section presents \emph{GARDRec}, a graph-grounded adaptive reasoning and decision-aware recommendation framework for LLM-based next-item ranking. The central design principle is to use the knowledge graph not merely as retrieved prompt evidence, but as a decision-level grounding layer that participates in item representation, user-context expansion, candidate comparison, and final ranking.

\subsection{Problem Formulation}

We consider the next-item ranking task under a multi-candidate evaluation setting. Let $\mathcal{U}=\{u_1,\dots,u_{|\mathcal{U}|}\}$ denote the user set and $\mathcal{V}=\{v_1,\dots,v_{|\mathcal{V}|}\}$ denote the item set. For a user $u$, the historical interaction sequence is represented as $\mathcal{H}_u=[v_1^u,\dots,v_{|\mathcal{H}_u|}^u]$, where $|\mathcal{H}_u|$ is the sequence length. 
Given a candidate set $\mathcal{C}_u=\{c_j\}_{j=1}^{M}$ with one ground-truth positive item and $M-1$ negative items, the model predicts the index $y \in \{1,\dots,M\}$ of the target item. Each candidate $c_j$ is also associated with a candidate label token $\ell_j$, such as option tokens A to T, which is later used by the restricted generative branch. To provide a fixed input shape, the user history is truncated or padded to length $N$.

To ground recommendation decisions in explicit relational knowledge, we introduce a structured knowledge graph $\mathcal{G}=(\mathcal{V}_g,\mathcal{E}_g)$. Each edge is associated with a semantic triplet $(h,r,t)$, where $h,t\in \mathcal{V}_g$ denote the head and tail entities and $r$ denotes their relation type. Items are linked to graph entities through their attributes, metadata, or external alignments. A graph representation learner maps each item-related entity $v$ into a dense embedding $e_v\in\mathbb{R}^{D_g}$, where $D_g$ is the graph embedding dimension. For each training instance, the historical items and candidate items are therefore represented as graph-aware matrices $H_{\text{emb}}\in\mathbb{R}^{N\times D_g}$ and $C_{\text{emb}}\in\mathbb{R}^{M\times D_g}$, which provide structured priors for subsequent context construction and ranking.

\subsection{Overall Framework}

\begin{figure}
    \centering
    \includegraphics[width=0.9\textwidth]{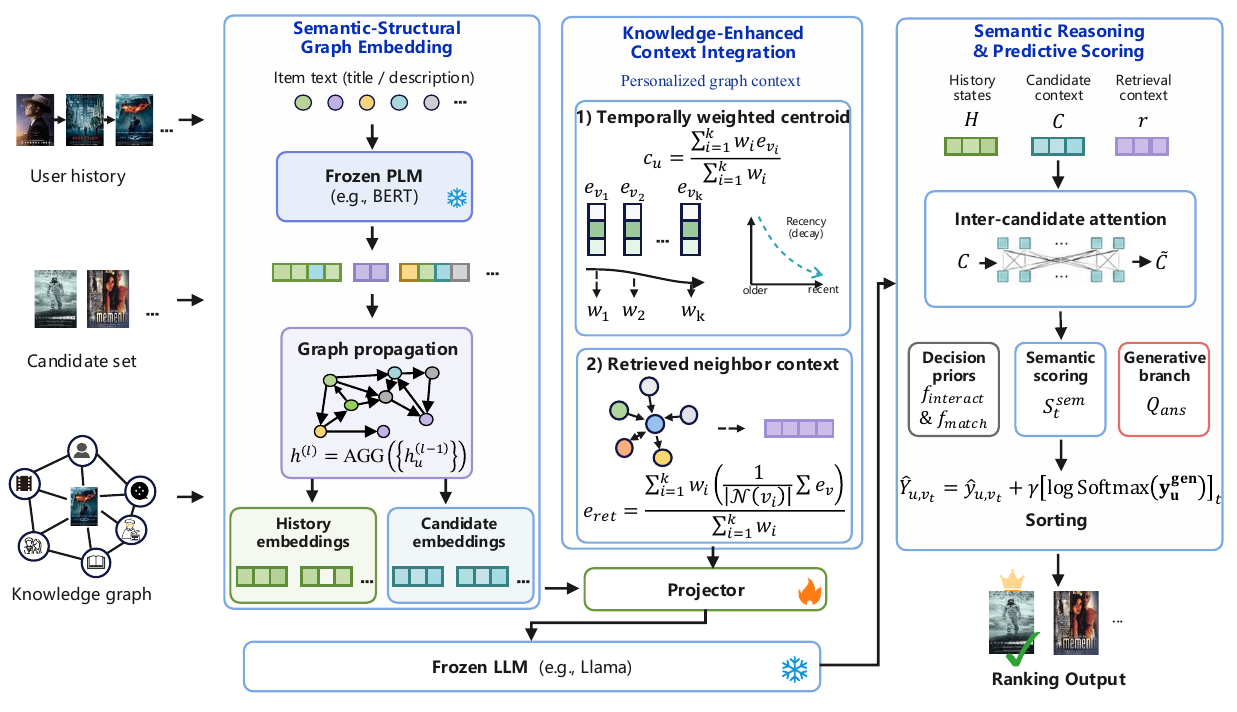}
    \caption{The overall framework of the proposed GARDRec model.}
    \label{fig:gardrec_framework}
    \vspace{-2mm}
\end{figure}

As illustrated in Figure~\ref{fig:gardrec_framework}, GARDRec consists of four coupled components: semantic-structural graph embedding, knowledge-enhanced context integration, multimodal prompt construction, and decision-aware predictive scoring. The first component builds item representations that combine textual semantics with graph topology. The second component converts user histories and graph neighborhoods into personalized contexts and explicit interaction features. The third component aligns graph-derived representations with the hidden space of a frozen LLM through projection, position encodings, slot embeddings, and soft markers. The fourth component extracts LLM hidden states, models inter-candidate competition, and combines semantic matching scores with explicit graph and collaborative priors.

This architecture separates two kinds of information that are often mixed in prompt-level RAG systems. Semantic and structural contexts are injected into the LLM as continuous prompts, allowing the LLM to perform contextual reasoning. Precise numerical signals, such as similarity, transition, co-occurrence, popularity, and hard attribute matching, bypass the text sequence and enter a late-stage scoring module. In this way, GARDRec preserves both the semantic flexibility of LLMs and the decision-level constraints required by multi-candidate recommendation.

\subsection{Semantic-Structural Graph Embedding}

The first module constructs graph embeddings that encode both node-level semantics and graph-level connectivity. Items and their attributes naturally form a knowledge graph $\mathcal{G}=(\mathcal{V}_g,\mathcal{E}_g)$, where nodes may correspond to items, categories, genres, authors, actors, or other domain-specific entities. Since many nodes contain textual descriptions or attribute names, we first use a frozen pretrained language model (PLM) to initialize each node representation:

\begin{equation}
x_v = \text{PLM}(\text{Text}_v) \in \mathbb{R}^{d_{plm}}\label{plm_text}
\end{equation}
where $\text{Text}_v$ denotes the raw textual description of node $v$, and $d_{plm}$ is the hidden dimension of the PLM output. This initialization provides a semantic representation before graph propagation, which is especially useful for sparse or cold-start items whose interaction histories are limited.

To integrate textual semantics with local topology, GARDRec applies GraphSAGE-style inductive propagation over $\mathcal{G}$. For any node $v\in\mathcal{V}_g$, the initial hidden state is $h_v^{(0)}=x_v$. At the $l$-th propagation layer, the model aggregates representations from the neighborhood set $\mathcal{N}(v)$:

\begin{equation}
    h_{\mathcal{N}(v)}^{(l)} = \text{AGGREGATE}^{(l)} \left( \left\{ h_u^{(l-1)} \mid u \in \mathcal{N}(v) \right \} \right)\label{arr_neig}
\end{equation}

The node representation is then updated by combining its previous state with the aggregated neighborhood representation:

\begin{equation}
    h_v^{(l)} = \sigma \left( W^{(l)} \left[ h_v^{(l-1)} \parallel h_{\mathcal{N}(v)}^{(l)} \right] \right)\label{concat_trans}
\end{equation}
where $W^{(l)}$ is the learnable transformation matrix, $\parallel$ denotes vector concatenation, $\sigma(\cdot)$ is a non-linear activation function, and $\text{AGGREGATE}^{(l)}$ is a permutation-invariant mean aggregator. 
After $L$ propagation layers, the final node representation is denoted as $e_v=h_v^{(L)}\in\mathbb{R}^{d_{graph}}$. These embeddings are stored as graph-aware item representations and reused by the context-integration, prompt-construction, and scoring modules.

\subsection{Knowledge-Enhanced Context Integration}

Static graph embeddings encode global semantic and structural properties, but next-item ranking also requires a dynamic user-specific context. GARDRec therefore constructs two forms of personalized graph context: a temporally weighted historical centroid and a retrieval-augmented neighborhood representation.

Given the recent historical sequence $\mathcal{H}_u=[v_1,v_2,\dots,v_k]$, where $v_k$ is the most recent item, we assign each historical item an exponential recency weight $w_i=\exp(i-k)$. The temporally weighted user centroid is computed as

\begin{equation}
    c_u = \frac{\sum_{i=1}^{k} w_i e_{v_i}}{\sum_{i=1}^{k} w_i}
\end{equation}
The centroid $c_u$ summarizes the user's dominant short-term preference in the graph embedding space. To move beyond the limited receptive field of observed user behavior, GARDRec further retrieves the first-order graph neighborhood $\mathcal{N}(v_i)$ of each historical item. The retrieved neighborhood embeddings are pooled and weighted by the same temporal decay factors:

\begin{equation}
    e_{ret} = \frac{\sum_{i=1}^{k} w_i \left( \frac{1}{|\mathcal{N}(v_i)|} \sum_{v \in \mathcal{N}(v_i)} e_v \right)}{\sum_{i=1}^{k} w_i}
\end{equation}
The resulting representation $e_{ret}$ serves as an external graph-grounded context. It expands the user's observed history with related entities and attributes, while still respecting the temporal importance of recent interactions.

For each candidate item $v_t$, GARDRec also extracts explicit decision features that are difficult to preserve through textual prompting alone. The interaction feature $f_{interact}$ concatenates graph similarity, sequential transition, co-occurrence, and statistical prior signals:

\begin{equation}
    f_{interact} = \left[ \text{Sim}(v_t, \mathcal{H}_u) \parallel \text{Trans}(v_t, \mathcal{H}_u) \parallel \text{Co-occur}(v_t, \mathcal{H}_u) \parallel \text{Stats}(u, v_t) \right] \in \mathbb{R}^{d_{interact}}
\end{equation}
Here, $\text{Sim}(v_t,\mathcal{H}_u)$ contains the cosine similarity between $e_{v_t}$ and the temporal centroid $c_u$, together with the maximum cosine similarity between $e_{v_t}$ and historical item embeddings. $\text{Trans}(v_t,\mathcal{H}_u)$ is computed from normalized first-order transition statistics between the most recent historical items and the candidate item. $\text{Co-occur}(v_t,\mathcal{H}_u)$ records normalized collaborative co-occurrence frequencies between the candidate and historical items in the training data. $\text{Stats}(u,v_t)$ includes scalar priors such as user activity, item popularity, and candidate retrieval rank statistics. All scalar features are normalized before being fed into the scoring MLP.
In parallel, the hard-matching feature is defined as

\begin{equation}
    f_{match} = \left[ \text{Attr}(v_t) \parallel \text{Rank}(v_t) \right] \in \mathbb{R}^{d_{match}}
\end{equation}
where $\text{Attr}(v_t)$ measures normalized overlap between candidate attributes and the attributes associated with the user's historical items, and $\text{Rank}(v_t)$ denotes the retrieval-stage ranking prior. Both $f_{interact}$ and $f_{match}$ are used as late-stage decision constraints rather than being serialized into the prompt. This bypass design reduces the risk that precise numerical evidence is diluted by deep self-attention over long multimodal sequences.

\subsection{Multimodal Prompt Construction}

LLMs natively operate on token embeddings, whereas GARDRec uses graph embeddings and dense retrieval contexts. To bridge this modality gap, we construct a continuous multimodal prompt that aligns graph-grounded representations with the LLM hidden space.

The natural-language task instruction $T_{text}$ is first mapped into token embeddings:

\begin{equation}
    E_{text} = \text{Embed}(T_{text})
\end{equation}

Then, a learnable non-linear projection network $\text{Proj}(\cdot)$ maps graph embeddings into the LLM hidden dimension $d_{LLM}$. To preserve the order of historical behaviors and distinguish candidate slots, GARDRec adds absolute position embeddings $p_i$ to historical nodes and candidate slot embeddings $s_j$ to candidate nodes. The projected representations are defined as

\begin{equation}
    h_{v_i} = \text{Proj}(e_{v_i}) + p_i, \quad h_{c_j} = \text{Proj}(e_{c_j}) + s_j, \quad h_{ret} = \text{Proj}(e_{ret})
\end{equation}

The historical matrix $E_{hist}$ is formed by stacking $\{h_{v_i}\}_{i=1}^{N}$, and the candidate matrix $E_{cand}$ is formed by stacking $\{h_{c_j}\}_{j=1}^{M}$. We further introduce learnable soft markers, including the history marker $M_{hist}$, retrieval marker $M_{ret}$, and candidate marker $M_{cand}$, to separate different context segments. The final multimodal prompt is constructed as

\begin{equation}
    E_{prompt} = \left[ E_{text} \parallel M_{hist} \parallel E_{hist} \parallel M_{ret} \parallel h_{ret} \parallel M_{cand} \parallel E_{cand} \parallel E_{ans} \parallel Q_{ans} \right]
\end{equation}
Here, $E_{ans}$ denotes a short answer-instruction segment that indicates the model should select one candidate label from the candidate set. $Q_{ans}$ is a learnable answer query token whose final hidden state is used by the restricted generative branch to predict candidate label tokens. The complete sequence $E_{prompt}$ is fed into the frozen LLM backbone for causal self-attention. The explicit interaction features $f_{interact}$ and $f_{match}$ are deliberately excluded from this prompt and are instead passed to the scoring module.

\subsection{Semantic Reasoning and Predictive Scoring}

After the LLM encodes the multimodal prompt, GARDRec extracts the final-layer hidden states corresponding to the historical items, retrieval context, and candidate items. We denote them as $H\in\mathbb{R}^{N\times d}$, $r\in\mathbb{R}^{d}$, and $C\in\mathbb{R}^{M\times d}$, respectively.

Because autoregressive self-attention processes the prompt under a causal mask, candidate representations may not fully capture mutual competition among candidates. GARDRec therefore applies a bidirectional inter-candidate attention layer over the candidate sequence:

\begin{equation}
    \tilde{C} = \text{LayerNorm}\left( C + \text{MHSA}(C, C, C) \right)
\end{equation}
where $\text{MHSA}(\cdot)$ denotes multi-head self-attention, and $\tilde{c}_t$ is the refined representation of the $t$-th candidate. This layer enables each candidate to be scored with awareness of the other candidates in the same ranking instance.

To model high-order matching between candidates and user contexts, we define a feature crossing operator $\Phi(\cdot,\cdot)$. For two representations $x$ and $y$, the operator is
\begin{equation}
    \Phi(x, y) = [x \parallel y \parallel x \odot y \parallel x - y]
\end{equation}
where $\odot$ denotes the Hadamard product. For a candidate $\tilde{c}_t$, GARDRec computes a semantic score $S_t^{sem}$ through three pathways. The attention pathway computes pairwise matching scores $s_{t,i}^{pair}=\text{MLP}_{hist}(\Phi(\tilde{c}_t,h_i))$ between the candidate and each historical state $h_i$, and obtains attention weights $\alpha_{t,i}$ by applying a softmax over historical positions. The recency pathway builds a temporally weighted historical representation $\bar{h}=\sum_{i=1}^{N}\beta_i h_i$, where $\beta_i=\frac{\exp(i-N)}{\sum_{j=1}^{N}\exp(j-N)}$. The retrieval pathway independently measures the compatibility between the candidate and the retrieved graph context $r$. These pathways are aggregated as
\begin{equation}
    S_{t}^{sem} = \sum_{i=1}^N \alpha_{t,i} s_{t,i}^{pair} + \eta_1 \cdot \text{MLP}_{mean}(\Phi(\tilde{c}_t, \bar{h})) + \eta_2 \cdot \text{MLP}_{ret}(\Phi(\tilde{c}_t, r))
\end{equation}
where $\eta_1$ and $\eta_2$ are fixed balancing weights.

The semantic score is then calibrated by explicit graph and collaborative priors. The hard-matching feature $f_{match}$ and interaction feature $f_{interact}$ are processed by independent Layer Normalization and MLP branches. To avoid unstable early training caused by noisy interaction features, the interaction branch is weighted by a non-negative learnable coefficient through $\text{Softplus}(w_{inter})$. The discriminative ranking logit is

\begin{equation}
    \hat{y}_{u, v_t} = S_{t}^{sem} + 0.3 \cdot \text{MLP}_{match}(\text{LN}(f_{match})) + \text{Softplus}(w_{inter}) \cdot \text{MLP}_{inter}(\text{LN}(f_{interact}))
\end{equation}

In parallel, GARDRec maintains a lightweight generative branch through the answer query token $Q_{ans}$. Let $q_{ans}$ denote the final-layer hidden state corresponding to $Q_{ans}$. The hidden state $q_{ans}$ is fed into the language-model head, and the logits of candidate label tokens are extracted as the generative likelihood scores:
\begin{equation}
    y^{gen}_{u, v_t} = \text{LM\_Head}(q_{ans})[\ell_t]
\end{equation}
where $\ell_t$ is the label token associated with candidate $v_t$, such as an option token from A to T. Restricting the generative decision to candidate labels avoids unnecessary full-vocabulary competition and stabilizes the auxiliary generation objective.

During inference, GARDRec combines the discriminative ranking logit and the log-normalized generative likelihood. Let $\mathbf{y}^{gen}_{u}=[y^{gen}_{u,v_1},\dots,y^{gen}_{u,v_M}]$ denote the generative logits over all candidates. The final score for candidate $v_t$ is computed as
\begin{equation}
    {\hat{Y}}_{u, v_t} = {\hat{y}}_{u, v_t} + \gamma \left[
    \log \operatorname{Softmax}(\mathbf{y}^{gen}_{u})
    \right]_t
\end{equation}
where $\gamma$ controls the contribution of the generative branch. The final candidate ranking is obtained by sorting candidates according to $\hat{Y}_{u,v_t}$.

\subsection{Joint Model Optimization}

GARDRec is trained with a multi-task objective that balances discriminative ranking and generative alignment. For candidate classification, the model uses a smoothed cross-entropy loss over the final candidate scores $\hat{Y}$:
\begin{equation}
    \mathcal{L}_{CE} = - \sum_{i=1}^{M} q_i \log p_i
\end{equation}
where $M$ is the number of candidates, $p_i=\text{Softmax}(\hat{Y})_i$ is the predicted probability of the $i$-th candidate, and $q_i$ is the smoothed target distribution. Label smoothing reduces overconfidence and improves the calibration of the candidate distribution.

To enlarge the margin between positive and hard negative candidates, we further introduce a Bayesian Personalized Ranking (BPR) loss over hard negative pairs:
\begin{equation}
    \mathcal{L}_{BPR} = - \sum_{\mathclap{(u, v_i, v_j) \in \mathcal{O}_{hard}}} \ln \sigma(\hat{y}_{u, v_i} - \hat{y}_{u, v_j})
\end{equation}
where $\mathcal{O}_{hard}$ denotes the set of pairwise observations selected by hard negative mining, $\sigma(\cdot)$ is the sigmoid function, $v_i$ is the positive candidate, and $v_j$ is a negative candidate with a relatively high predicted score in the same candidate set.

For the auxiliary generative objective, GARDRec restricts the language-model probability space to the candidate label set $\mathcal{V}_{cand}$, such as option tokens A to T. The restricted likelihood loss is

\begin{equation}
    \mathcal{L}_{LM} = - \log \frac{\exp(s_{y})}{\sum_{c \in \mathcal{V}_{cand}} \exp(s_{c})}
\end{equation}
where $s_y$ is the logit of the correct candidate label and $s_c$ is the logit of candidate label $c$. This single-step objective keeps the generative branch aligned with valid recommendation options instead of optimizing over the full vocabulary.

The final training objective is a weighted sum of the three losses:

\begin{equation}
    \mathcal{L} = \mathcal{L}_{CE} + \lambda_1 \mathcal{L}_{BPR} + \lambda_2 \mathcal{L}_{LM}
\end{equation}
where $\lambda_1$ and $\lambda_2$ balance the pairwise ranking and generative objectives. This optimization scheme encourages GARDRec to retain LLM-based semantic reasoning while learning graph-grounded ranking boundaries and explicit collaborative constraints.

\subsection{Complexity Analysis}

We analyze the complexity of GARDRec by separating the offline graph representation stage from the online candidate-ranking stage. Let $n_g=|\mathcal{V}_g|$ and $m_g=|\mathcal{E}_g|$ denote the number of nodes and edges in the knowledge graph. Let $L_g$ be the number of graph propagation layers, $d_{graph}$ be the graph embedding dimension, $d$ be the LLM hidden dimension, $N$ be the maximum history length, $M$ be the number of candidates, and $S$ be the multimodal prompt length.

The semantic-structural graph representations are constructed offline and cached before recommendation inference. Initializing node features with a frozen text encoder costs $O(n_g C_{\mathrm{PLM}})$, where $C_{\mathrm{PLM}}$ is the average cost of encoding one node description. Given the initialized node features, the GraphSAGE-style propagation with mean aggregation costs
\begin{equation}
    O\left(L_g \left(m_g d_{graph} + n_g d_{graph}^{2}\right)\right)
\end{equation}
where the two terms correspond to edge-level neighborhood aggregation and node-level linear transformation, respectively. Since this stage is independent of a specific user query, the resulting graph-aware item embeddings can be reused across training and inference instances.

For each online ranking instance, GARDRec retrieves the embeddings of $N$ historical items and $M$ candidate items, and then constructs personalized graph contexts and explicit decision features. If $\bar{K}$ denotes the average number of first-order neighbors retrieved for each historical item, the temporal centroid and neighborhood pooling cost
\begin{equation}
    O\left((N+N\bar{K})d_{graph}\right)
\end{equation}
Candidate-wise interaction and matching feature construction costs
\begin{equation}
    O\left(MN d_{graph} + MN + M(d_{interact}+d_{match})\right)
\end{equation}
where $MN d_{graph}$ upper-bounds candidate--history similarity computation, and the lookup-based transition and co-occurrence statistics contribute the $MN$ term.

The multimodal prompt construction projects graph-derived vectors into the LLM hidden space. For $N$ history vectors, $M$ candidate vectors, and one retrieved graph context vector, the projection cost is
\begin{equation}
    O\left((N+M+1)d_{graph}d\right)
\end{equation}
During inference, the dominant online cost is the forward pass of the frozen LLM. With $L_{llm}$ transformer layers and prompt length $S$, the standard self-attention and feed-forward computation costs
\begin{equation}
    O\left(L_{llm}(S^{2}d + S d^{2})\right)
\end{equation}
During training, the LLM parameters remain frozen, while gradients are propagated through the LLM computation graph to update the projection network, soft markers, inter-candidate attention layer, and scoring modules. This keeps the same asymptotic order but increases the constant factor compared with inference.

After LLM encoding, the bidirectional inter-candidate attention layer introduces an additional $O(M^{2}d)$ cost. The semantic scoring pathways compare each candidate with historical states and the retrieved graph context, leading to $O(MNd+Md)$ operations. The late-stage matching and interaction branches add $O(M(d_{match}+d_{interact}))$ when the hidden dimensions of the small MLPs are treated as constant factors. If candidate-label logits are computed by restricted output projection, the generative scoring cost is $O(Md)$; otherwise, evaluating the full LM head first would introduce an additional $O(|\mathcal{V}_{tok}|d)$ cost, where $|\mathcal{V}_{tok}|$ is the vocabulary size. The cross-entropy loss costs $O(M)$, and the BPR loss costs $O(|\mathcal{O}_{hard}|)$ for the selected hard negative pairs.

Combining the online components, the per-instance ranking complexity is
\begin{equation}
\begin{aligned}
O(&L_{llm}(S^{2}d + S d^{2})
  + (N+M)d_{graph}d
  + M^{2}d  \\
  &+ MNd
  + MN d_{graph}
  + M(d_{interact}+d_{match})
  + |\mathcal{O}_{hard}|)
\end{aligned}
\end{equation}
In the candidate-ranking setting, $N$ and $M$ are fixed to small values, while graph embeddings are precomputed and cached. Therefore, the additional graph-grounded decision modules introduce limited overhead compared with the LLM encoding cost. The only quadratic term introduced by GARDRec itself is the inter-candidate attention cost $O(M^{2}d)$, which is manageable for reranking because $M$ is much smaller than the full item catalog size. Restricted candidate-label scoring further keeps the generative decision cost proportional to the candidate set size rather than the full vocabulary size.

\section{Experiments}

This section evaluates whether \emph{GARDRec} improves LLM-based next-item ranking by using graph knowledge as a decision-level grounding layer. We organize the experiments around four questions: (1) whether GARDRec outperforms representative graph-augmented LLM baselines under the same candidate-ranking protocol; (2) which components contribute to the improvement; (3) whether the gain remains under sparse-history and long-tail diagnostic settings compared with the closest KG-RAG baseline; and (4) how the method behaves under the boundary condition of strict zero-history cold-start recommendation.

\subsection{Experimental Setup} 

\subsubsection{Datasets and KG Construction} 

We conduct experiments on three public recommendation benchmarks: MovieLens-1M (ML-1M)\footnote{\url{https://grouplens.org/datasets/movielens/1m/}}, MovieLens-20M (ML-20M)\footnote{\url{https://grouplens.org/datasets/movielens/20m/}}, and Amazon-Book\footnote{\url{https://jmcauley.ucsd.edu/data/amazon/}}. ML-1M contains approximately one million movie ratings from 6,000 users over 4,000 movies, while ML-20M contains approximately 20 million interactions from 138,000 users over 27,000 movies. Amazon-Book is a much sparser e-commerce dataset with user--book interactions and product metadata. 

For the two movie datasets, item entities are aligned with TMDB metadata to construct graph attributes such as directors, actors, genres, and release years. For Amazon-Book, graph attributes such as authors, categories, and publication information are extracted from product metadata. Items and their associated attributes are organized as heterogeneous item knowledge graphs, where edges connect items with metadata entities through typed relations. These datasets cover both dense movie recommendation and sparse book recommendation, allowing us to evaluate whether GARDRec can exploit semantic metadata, graph neighborhoods, and collaborative interaction signals under different data conditions. Table~\ref{tab:dataset_statistics} reports the dataset and knowledge-graph statistics after preprocessing.

\begin{table}[t]
\centering
\caption{Statistics of the recommendation datasets and constructed item knowledge graphs.}
\label{tab:dataset_statistics}
\small
\setlength{\tabcolsep}{4pt}
\renewcommand{\arraystretch}{1.08}
\begin{tabular}{lrrrrrr}
\toprule
Dataset & Users & Items & Interactions & Entities & Relations & KG Triples \\
\midrule
ML-1M        & 6,038     & 3,533     & 575,281    & 250,631   & 264 & 348,979 \\
ML-20M       & 138,287   & 20,720    & 9,995,410  & 1,278,544 & 436 & 1,827,361 \\
Amazon-Book  & 6,106,019 & 1,891,460 & 13,886,788 & 186,954   & 16  & 259,861 \\
\bottomrule
\end{tabular}
\end{table}

\subsubsection{Evaluation Protocol and Metrics} 

We follow the leave-one-out evaluation protocol. For each user, the most recent interaction is used as the test target, and the preceding interactions are used as historical context. Following the multi-candidate ranking setting, each test instance contains one ground-truth item and 19 randomly sampled unobserved items, resulting in $M=20$ candidate items. All methods under the same dataset--backbone setting are evaluated with the same candidate sets. Unless otherwise specified, the input history is truncated or padded to $N=10$ items.

We report Accuracy (ACC), Recall@3 (R@3), and Recall@5 (R@5). In this candidate-ranking protocol, ACC is equivalent to HR@1 and measures whether the ground-truth item is ranked first. R@3 and R@5 measure whether the ground-truth item appears in the top-3 and top-5 ranked candidate lists, respectively. For the long-tail diagnostic analysis, we additionally report NDCG@5 to evaluate the discounted rank of the target item within the top-ranked candidates.

\subsubsection{Baselines} 

We compare GARDRec with representative graph-augmented LLM baselines under three backbone LLMs: LLaMA2-7B, LLaMA3-8B, and Qwen2-7B. The baselines include two inference-only methods, KG-Text \citep{wuRetrieveRewriteAnswerKGtoTextEnhanced2023a} and KAPING \citep{baekKnowledgeAugmentedLanguageModel2023a}, which serialize graph knowledge into textual prompts without training task-specific parameters. We also include frozen-LLM prompt-tuning baselines: PT w/ KG-Text, GraphToken w/ RAG \citep{perozziLetYourGraph2024a}, G-Retriever \citep{heGRetrieverRetrievalAugmentedGeneration2024a}, and K-RagRec \citep{wangKnowledgeGraphRetrievalAugmented2025}. These baselines represent prompt-level, graph-token-level, and knowledge-graph retrieval strategies for injecting graph evidence into LLM inference. 

The baseline design is aligned with the central question of this paper. Inference-only baselines test whether serialized graph evidence can be directly used by LLMs without task-specific training. Frozen-LLM prompt-tuning and graph-token baselines test whether continuous graph representations and retrieved graph contexts improve LLM-based recommendation while keeping the backbone frozen. GARDRec differs from these methods by using graph-derived representations and personalized graph contexts for LLM reasoning while preserving explicit interaction and matching features in a late-stage decision module. Therefore, the comparison focuses on graph-augmented LLM recommendation strategies rather than claiming coverage of every classical, sequential, or non-LLM recommender family. 

\subsubsection{Implementation Details}

All experiments are conducted on a single NVIDIA RTX 4090 GPU. Initial node text features are extracted with Sentence-BERT. The graph representation module uses a two-layer GraphSAGE encoder with output dimension $d_{graph}=128$. The LLM backbone is frozen, and the maximum input length is set to 512. For graph-context construction, GARDRec retrieves first-order graph neighborhoods associated with historical items and candidate items, and converts them into personalized graph context representations. For graph-retrieval baselines, retrieved graph evidence is formatted according to their original prompting or graph-token design under the same candidate-ranking protocol. All transition, co-occurrence, popularity, and retrieval-prior features are computed only from the training interactions and metadata, without using test targets.

The learnable components include the graph-to-LLM projection network, position embeddings, candidate slot embeddings, soft markers, inter-candidate attention, and scoring MLPs. The MLP for the interaction feature $f_{interact}$ contains two hidden layers. We optimize the model with AdamW using a learning rate of $1\times10^{-4}$ and a batch size of 16. The loss weights are set to $\lambda_1=0.3$ for the BPR loss and $\lambda_2=0.1$ for the restricted language-model loss.

Freezing the LLM backbone keeps the comparison focused on how graph evidence is represented, aligned, and used for ranking, rather than on full LLM fine-tuning. The learnable projection and soft prompt components adapt graph embeddings to the LLM hidden space, while the scoring branches preserve structured features that are not naturally expressed as text.

\subsection{Overall Performance}

Table~\ref{tab:overall_results} reports the main comparison across three datasets and three LLM backbones, with K-RagRec included as the closest KG-RAG recommendation baseline. Under this stricter comparison, GARDRec achieves the best result in 24 of the 27 reported metric columns. The exceptions are LLaMA3-8B on ML-1M for ACC, where K-RagRec is marginally higher (0.472 vs. 0.471), and Qwen2-7B on ML-20M for R@3 and R@5, where K-RagRec obtains slightly higher recall. This pattern supports a more precise claim: GARDRec usually improves both top-1 ranking and top-$K$ retention, while prompt-level KG-RAG can remain competitive in a few backbone--dataset combinations.

\begin{table*}[t]
\centering
\caption{Performance comparison across backbone LLMs and datasets. The best results are highlighted in bold. Relative change is computed against the strongest non-GARDRec baseline under the same backbone, dataset, and metric.}
\label{tab:overall_results}
\footnotesize
\setlength{\tabcolsep}{3.4pt}
\renewcommand{\arraystretch}{1.12}

\begin{tabular*}{\textwidth}{@{\extracolsep{\fill}}llccccccccc@{}}
\toprule
\multirow{2}{*}{Setting} 
& \multirow{2}{*}{Method} 
& \multicolumn{3}{c}{ML-1M} 
& \multicolumn{3}{c}{ML-20M} 
& \multicolumn{3}{c}{Amazon-Book} \\
\cmidrule(lr){3-5} \cmidrule(lr){6-8} \cmidrule(lr){9-11}
& & ACC & R@3 & R@5 & ACC & R@3 & R@5 & ACC & R@3 & R@5 \\
\midrule

\multicolumn{11}{l}{\textbf{LLaMA2-7B}} \\
Infer. & KG-Text       & 0.076 & --    & --    & 0.052 & --    & --    & 0.058 & --    & --    \\
Infer. & KAPING        & 0.079 & --    & --    & 0.069 & --    & --    & 0.063 & --    & --    \\
PT     & PT w/ KG-Text & 0.078 & 0.191 & 0.308 & 0.051 & 0.152 & 0.250 & 0.074 & 0.165 & 0.245 \\
PT     & GraphToken    & 0.268 & 0.421 & 0.466 & 0.186 & 0.433 & 0.576 & 0.326 & 0.515 & 0.624 \\
PT     & G-Retriever   & 0.274 & 0.532 & 0.650 & 0.342 & 0.619 & 0.739 & 0.275 & 0.487 & 0.612 \\
PT     & K-RagRec      & 0.435 & 0.725 & 0.831 & 0.600 & 0.850 & 0.913 & 0.508 & 0.690 & 0.780 \\
PT     & GARDRec       & \textbf{0.468} & \textbf{0.792} & \textbf{0.909} 
                         & \textbf{0.641} & \textbf{0.872} & \textbf{0.917} 
                         & \textbf{0.585} & \textbf{0.785} & \textbf{0.851} \\
\multicolumn{2}{r}{\textit{Rel. change}} 
       & \textit{7.6\%}  & \textit{9.2\%}  & \textit{9.4\%} 
       & \textit{6.8\%}  & \textit{2.6\%}  & \textit{0.4\%} 
       & \textit{15.2\%} & \textit{13.8\%} & \textit{9.1\%} \\

\midrule

\multicolumn{11}{l}{\textbf{LLaMA3-8B}} \\
Infer. & KG-Text       & 0.095 & --    & --    & 0.060 & --    & --    & 0.054 & --    & --    \\
Infer. & KAPING        & 0.084 & --    & --    & 0.069 & --    & --    & 0.062 & --    & --    \\
PT     & PT w/ KG-Text & 0.134 & 0.294 & 0.433 & 0.094 & 0.205 & 0.296 & 0.083 & 0.207 & 0.314 \\
PT     & GraphToken    & 0.355 & 0.622 & 0.737 & 0.473 & 0.719 & 0.805 & 0.428 & 0.567 & 0.661 \\
PT     & G-Retriever   & 0.352 & 0.632 & 0.746 & 0.502 & 0.736 & 0.796 & 0.417 & 0.584 & 0.682 \\
PT     & K-RagRec      & \textbf{0.472} & 0.704 & 0.765 & 0.634 & 0.779 & 0.818 & 0.514 & 0.662 & 0.723 \\
PT     & GARDRec       & 0.471 & \textbf{0.813} & \textbf{0.929} 
                         & \textbf{0.647} & \textbf{0.868} & \textbf{0.927} 
                         & \textbf{0.572} & \textbf{0.782} & \textbf{0.842} \\
\multicolumn{2}{r}{\textit{Rel. change}} 
       & \textit{-0.2\%} & \textit{15.5\%} & \textit{21.4\%} 
       & \textit{2.1\%}  & \textit{11.4\%} & \textit{13.3\%} 
       & \textit{11.3\%} & \textit{18.1\%} & \textit{16.5\%} \\

\midrule

\multicolumn{11}{l}{\textbf{Qwen2-7B}} \\
Infer. & KG-Text       & 0.160 & --    & --    & 0.174 & --    & --    & 0.194 & --    & --    \\
Infer. & KAPING        & 0.196 & --    & --    & 0.208 & --    & --    & 0.220 & --    & --    \\
PT     & PT w/ KG-Text & 0.190 & 0.371 & 0.499 & 0.259 & 0.397 & 0.494 & 0.303 & 0.451 & 0.553 \\
PT     & GraphToken    & 0.259 & 0.487 & 0.608 & 0.370 & 0.550 & 0.632 & 0.365 & 0.568 & 0.658 \\
PT     & G-Retriever   & 0.304 & 0.551 & 0.644 & 0.389 & 0.606 & 0.685 & 0.355 & 0.552 & 0.658 \\
PT     & K-RagRec      & 0.416 & 0.712 & 0.829 & 0.586 & \textbf{0.842} & \textbf{0.904} & 0.502 & 0.686 & 0.767 \\
PT     & GARDRec       & \textbf{0.444} & \textbf{0.799} & \textbf{0.925} 
                         & \textbf{0.623} & 0.839 & 0.884 
                         & \textbf{0.508} & \textbf{0.731} & \textbf{0.829} \\
\multicolumn{2}{r}{\textit{Rel. change}} 
       & \textit{6.7\%}  & \textit{12.2\%} & \textit{11.6\%} 
       & \textit{6.3\%}  & \textit{-0.4\%} & \textit{-2.2\%} 
       & \textit{1.2\%}  & \textit{6.6\%}  & \textit{8.1\%} \\

\bottomrule
\end{tabular*}

\vspace{2pt}
\begin{minipage}{0.98\textwidth}
\footnotesize 
\emph{Note:} ``Infer.'' denotes inference-only prompting, and ``PT'' denotes frozen-LLM prompt tuning. GraphToken refers to GraphToken w/ RAG. K-RagRec is included as the closest knowledge-graph RAG recommendation baseline. A dash indicates that top-$K$ ranking statistics are not reported for inference-only baselines.
\end{minipage}
\end{table*}

Several patterns can be observed from Table~\ref{tab:overall_results}. First, inference-only methods remain weak in this candidate-ranking protocol even when graph evidence is verbalized as text, suggesting that factual availability alone is insufficient for next-item ranking. Second, K-RagRec consistently outperforms earlier prompt-level and graph-token baselines, confirming that KG retrieval provides a strong reference point for graph-augmented LLM recommendation. Third, GARDRec outperforms the strongest non-GARDRec baseline in most settings, with relative changes ranging from $-0.2\%$ to 15.2\% for ACC, from $-0.4\%$ to 18.1\% for R@3, and from $-2.2\%$ to 21.4\% for R@5. The few negative entries are small and concentrated in settings where K-RagRec already provides strong candidate recall.

The comparison with K-RagRec further clarifies the effect of decision-level graph grounding. K-RagRec retrieves graph evidence to support LLM reasoning, whereas GARDRec additionally converts graph-derived representations, personalized contexts, interaction statistics, and generative likelihoods into an explicit late-stage ranking decision. This design leads to consistent gains over GraphToken and G-Retriever and stronger R@3/R@5 values in most K-RagRec comparisons. The smaller but still positive gains on Amazon-Book under Qwen2-7B suggest that item-side textual and metadata evidence can make KG-RAG prompting competitive. Overall, the main benchmark supports the central argument that graph evidence is more effective when it participates in representation learning, context construction, candidate comparison, and score calibration, rather than only enriching the prompt.

\subsection{Ablation Study}

Table~\ref{tab:ablation} reports a unified ablation study of GARDRec under the Qwen2-7B backbone on ML-1M. The full model corresponds to the Qwen2-7B/ML-1M result in Table~\ref{tab:overall_results}. The ablation includes both core component removals and stronger diagnostic simplifications that replace graph-encoded representations with text-only retrieval or remove multiple decision-level signals. Since these variants are evaluated in one representative backbone--dataset setting rather than repeated for every setting in Table~\ref{tab:overall_results}, they are used as module-level diagnostic evidence. Unless otherwise specified, the following sparse-history, long-tail, and strict zero-history diagnostic experiments are also conducted on ML-1M with Qwen2-7B.

\begin{table*}[t]
\centering
\caption{Unified ablation study of GARDRec with Qwen2-7B on ML-1M. The full model is aligned with the corresponding main-table result. Diagnostic simplifications remove broader functional blocks, while component removals isolate individual modules.}
\label{tab:ablation}
\footnotesize
\setlength{\tabcolsep}{4pt}
\renewcommand{\arraystretch}{1.08}
\begin{tabular*}{\textwidth}{@{\extracolsep{\fill}}llcccc@{}}
\toprule
Variant & Removed or simplified module & ACC & R@3 & R@5 & $\Delta$ ACC \\
\midrule
GARDRec (Full) & None & \textbf{0.444} & \textbf{0.799} & \textbf{0.925} & -- \\
\midrule
\multicolumn{6}{l}{\textit{Representation and context diagnostics}} \\
TextRAG diagnostic & GraphSAGE embeddings replaced by text retrieval & 0.128 & 0.280 & 0.380 & -71.2\% \\
w/o Graph Proj. & Graph-to-LLM projection & 0.154 & 0.334 & 0.435 & -65.3\% \\
w/o Retrieval-Aug. & Retrieved neighbor context & 0.332 & 0.610 & 0.752 & -25.2\% \\
\midrule
\multicolumn{6}{l}{\textit{Decision-level grounding diagnostics}} \\
w/o Decision Grounding & Late-fusion branches and BPR loss & 0.236 & 0.419 & 0.505 & -46.8\% \\
w/o Late Fusion & Late-stage scoring branches & 0.280 & 0.573 & 0.724 & -36.9\% \\
w/o Ranking Loss & Pairwise ranking constraint & 0.355 & 0.641 & 0.754 & -20.0\% \\
w/o Match Features & Hard matching features & 0.357 & 0.634 & 0.749 & -19.6\% \\
w/o Gen-Branch & Restricted generative branch & 0.361 & 0.642 & 0.779 & -18.7\% \\
\bottomrule
\end{tabular*}
\end{table*}

The most salient pattern in Table~\ref{tab:ablation} is the importance of graph representation and alignment. The TextRAG diagnostic replaces graph-encoded representations with retrieved textual graph evidence, reducing ACC from 0.444 to 0.128 and R@5 from 0.925 to 0.380. Removing the graph-to-LLM projection also leads to a severe ACC drop of 65.3\%. These results indicate that GARDRec benefits not only from exposing graph facts to the LLM, but also from transforming graph structure into semantic-structural representations that can be aligned with the frozen LLM space and used by the downstream ranking module.

Decision-level grounding forms the second group of influential factors. The broader ``w/o Decision Grounding'' diagnostic removes late-fusion branches together with the pairwise ranking constraint, reducing ACC by 46.8\%. Removing only the late-fusion branches still causes a 36.9\% ACC drop, while retaining a relatively higher R@5 of 0.724. This suggests that simplified scoring can still keep some relevant items in the top-ranked list, but is less effective at calibrating the final candidate order. The result supports the central design of GARDRec: graph-derived representations and retrieved contexts should be coupled with explicit decision-level scoring rather than only injected into the prompt.

The remaining component removals show that retrieval augmentation, ranking supervision, hard matching, and the generative branch provide complementary benefits. Removing retrieved neighbor context reduces ACC by 25.2\%, confirming the value of first-order graph neighborhoods as personalized context beyond the observed user sequence. Removing the pairwise ranking loss, hard matching features, and restricted generative branch yields ACC drops of 20.0\%, 19.6\%, and 18.7\%, respectively. In particular, removing the generative branch weakens both ACC and R@5, from 0.444 to 0.361 and from 0.925 to 0.779, suggesting that candidate-label likelihood helps calibrate the final ranking distribution in addition to the discriminative score.

\subsection{Diagnostic Analysis under Sparse and Long-tail Conditions}

The main benchmark evaluates candidate-level ranking performance on complete test sets. To further examine whether the gain over prompt-level KG-RAG is preserved in harder recommendation regimes, we conduct two diagnostic analyses with Qwen2-7B on ML-1M using fixed sampled subsets: user-history sparsity and item popularity. In each diagnostic table, GARDRec and K-RagRec are evaluated on the same instances and candidate sets. 

\subsubsection{User-history Sparsity}

We group diagnostic instances according to the number of available historical interactions before padding or truncation. To make the comparison balanced across history-length regimes, we use the same number of sampled instances for each group and evaluate GARDRec and K-RagRec on identical candidate sets.

\begin{table}[t]
\centering
\caption{User-history sparsity analysis with Qwen2-7B on ML-1M. History groups are formed according to the number of available historical interactions. Best results within each group are highlighted in bold.}
\label{tab:user_sparsity}
\small
\setlength{\tabcolsep}{5pt}
\renewcommand{\arraystretch}{1.08}
\begin{tabular}{llccc}
\toprule
History group & Method & Samples & ACC & R@5 \\
\midrule
\multirow{2}{*}{History 1--3} 
& GARDRec & 886 & \textbf{0.3815} & \textbf{0.8725} \\
& K-RagRec & 886 & 0.3093 & 0.7460 \\
\midrule
\multirow{2}{*}{History 4--6} 
& GARDRec & 886 & \textbf{0.4097} & \textbf{0.8840} \\
& K-RagRec & 886 & 0.4029 & 0.8296 \\
\midrule
\multirow{2}{*}{History 7--10} 
& GARDRec & 886 & 0.4199 & \textbf{0.8916} \\
& K-RagRec & 886 & \textbf{0.4436} & 0.8488 \\
\midrule
\multirow{2}{*}{Overall} 
& GARDRec & 2658 & \textbf{0.4037} & \textbf{0.8829} \\
& K-RagRec & 2658 & 0.3853 & 0.8081 \\
\bottomrule
\end{tabular}
\end{table}

Table~\ref{tab:user_sparsity} shows that the advantage of GARDRec is most pronounced when user histories are short. In the 1--3 interaction group, GARDRec improves ACC from 0.3093 to 0.3815 and R@5 from 0.7460 to 0.8725 over K-RagRec. This result is consistent with the motivation of graph-grounded recommendation: when only a few user actions are observed, converting sparse behavioral evidence into graph-structured preference context is more effective than relying only on prompt-level KG snippets.

The overall diagnostic comparison further supports this observation. Across 2658 history-length diagnostic instances, GARDRec improves ACC by 4.8\% and R@5 by 9.3\% relative to K-RagRec. The only exception is top-1 ACC in the 7--10 group, where K-RagRec obtains 0.4436 compared with 0.4199 for GARDRec. However, GARDRec still achieves a higher R@5 in this group. This suggests that when more user history is available, prompt-level KG-RAG can sometimes identify the target item at the first position, whereas GARDRec more consistently retains the target item within the top-ranked candidate set through graph-grounded user representation and candidate comparison.

\subsubsection{Long-tail Item Groups}

We further group diagnostic instances according to the popularity of the ground-truth item in the training set. Items are grouped according to their interaction frequency in the training set. Head, medium, and tail-cold groups are formed by descending popularity quantiles, with tail-cold items corresponding to the lowest-frequency group. GARDRec and K-RagRec are evaluated on the same instances and candidate sets within each group.

\begin{table}[t]
\centering
\caption{Long-tail item analysis with Qwen2-7B on ML-1M. GARDRec and K-RagRec are compared within the same diagnostic popularity groups. Best results within each group are highlighted in bold.}
\label{tab:long_tail}
\small
\setlength{\tabcolsep}{4pt}
\renewcommand{\arraystretch}{1.08}
\begin{tabular}{llcccc}
\toprule
Popularity group & Method & Samples & ACC & R@5 & NDCG@5 \\
\midrule
\multirow{2}{*}{Head}
& GARDRec & 435 & \textbf{0.5425} & \textbf{0.9885} & \textbf{0.7894} \\
& K-RagRec & 435 & 0.5011 & 0.9149 & 0.7323 \\
\midrule
\multirow{2}{*}{Medium}
& GARDRec & 371 & \textbf{0.3881} & \textbf{0.9488} & \textbf{0.6867} \\
& K-RagRec & 371 & 0.3801 & 0.7924 & 0.5981 \\
\midrule
\multirow{2}{*}{Tail-cold}
& GARDRec & 91 & 0.3187 & \textbf{0.9231} & \textbf{0.6315} \\
& K-RagRec & 91 & \textbf{0.3736} & 0.8241 & 0.6060 \\
\midrule
\multirow{2}{*}{Overall}
& GARDRec & 897 & \textbf{0.4559} & \textbf{0.9654} & \textbf{0.7309} \\
& K-RagRec & 897 & 0.4381 & 0.8551 & 0.6639 \\
\bottomrule
\end{tabular}
\end{table}

Table~\ref{tab:long_tail} examines whether decision-level graph grounding remains useful across different item popularity regimes. Overall, GARDRec outperforms K-RagRec by 4.1\% in ACC, 12.9\% in R@5, and 10.1\% in NDCG@5. The gains are especially clear in the head and medium groups, where GARDRec improves all three metrics. Since both methods use the same backbone and candidate sets, these improvements suggest that the advantage does not simply come from stronger language-model semantics or popularity bias. Instead, it comes from allowing graph representations, personalized graph context, and explicit decision features to participate in candidate ordering. This comparison further supports our central argument that graph evidence should not only be retrieved as prompt-level context, but should also contribute to candidate comparison and final score calibration.

The tail-cold group presents a more nuanced case. K-RagRec achieves higher ACC, while GARDRec achieves higher R@5 and NDCG@5. This indicates that prompt-level KG evidence can still be effective for placing a rare target item at the first position in some cases, especially when item-side descriptions or KG facts are highly discriminative. However, GARDRec retains the target item more consistently within the top-five list and assigns it a better discounted rank on average. This pattern is consistent with the role of decision-level graph grounding: even when exact top-1 prediction for the coldest items remains difficult, graph-grounded candidate comparison can improve the stability of the ranked list. Considering the smaller sample size of the tail-cold group, this result should be interpreted as diagnostic evidence rather than a definitive long-tail conclusion. Overall, the analysis supports a bounded claim: decision-level graph grounding improves candidate comparison across most popularity regimes, while exact top-1 prediction for the coldest items remains challenging when behavioral and relational evidence is limited.

\subsection{Strict Zero-history Cold-start Analysis}


The previous Qwen2-7B/ML-1M diagnostics still provide short user histories. We further evaluate a stricter zero-history setting, where the model receives no user interaction history and must rank candidates using only item-side graph metadata and textual attributes. This setting intentionally removes the history-conditioned graph context used by GARDRec, and therefore serves as a boundary test rather than a standard advantage claim.

\begin{table}[t]
\centering
\caption{Strict zero-history diagnostic evaluation with Qwen2-7B on ML-1M over 494 instances. K-RagRec performs better in this item-only setting, highlighting a boundary condition for GARDRec when user-history grounding is unavailable.}
\label{tab:cold_start}
\small
\setlength{\tabcolsep}{6pt}
\renewcommand{\arraystretch}{1.08}
\begin{tabular}{lccc}
\toprule
Method & Samples & ACC & R@5 \\
\midrule
GARDRec & 494 & 0.3421 & 0.7085 \\
K-RagRec & 494 & \textbf{0.4737} & \textbf{0.8826} \\
\bottomrule
\end{tabular}
\end{table}

Table~\ref{tab:cold_start} shows that K-RagRec is substantially stronger in the strict zero-history setting. This result does not contradict the sparse-history and long-tail findings; instead, it clarifies the operating condition of the proposed method. GARDRec is designed to transform limited user behavior into graph-grounded preference context and combine this context with candidate-side graph evidence. When user behavior is completely absent, its temporal centroid, history-conditioned neighborhood retrieval, and interaction-aware matching signals are largely disabled. In contrast, K-RagRec can directly verbalize item-side KG facts in the prompt, which is particularly suitable for item-only cold-start ranking.

This boundary result further clarifies the mechanism behind GARDRec. Decision-level graph grounding is most effective when graph evidence can be conditioned on at least limited user behavior. When no user history is available, the task is reduced to item-side evidence matching, where prompt-level KG verbalization can be more suitable. Therefore, the zero-history result does not weaken the main claim; instead, it shows that the advantage of GARDRec comes from transforming observed user behavior into graph-grounded preference context for candidate comparison and score calibration. In practical deployment, an item-only KG prompting strategy can be used as a fallback for users with no observed interactions, while GARDRec is better suited to sparse-but-not-empty recommendation histories.

\subsection{Discussion}

The experiments support GARDRec as a graph-grounded reranking framework for LLM-based recommendation, while also clarifying its effective operating conditions. The overall performance comparison shows that GARDRec generally improves over representative graph-augmented LLM baselines across different datasets and backbone models. The ablation study further confirms that the improvement comes from multiple complementary components, including semantic-structural graph representation, graph-to-LLM alignment, retrieved neighborhood context, decision-level scoring, pairwise ranking supervision, hard matching features, and the restricted generative branch.

The sparse-history and long-tail diagnostic analyses further explain when decision-level graph grounding is most useful. Under short user histories, GARDRec improves over K-RagRec by transforming limited behavioral evidence into graph-grounded preference context. Across item popularity groups, GARDRec generally achieves stronger top-$K$ ranking quality, indicating that graph-derived representations and explicit decision features help stabilize candidate ordering beyond prompt-level KG retrieval. The tail-cold group also reveals a more nuanced pattern: K-RagRec can obtain better top-1 accuracy for some rare items, while GARDRec better retains the target item within the top-ranked list. This suggests that decision-level graph grounding is particularly useful for candidate comparison and ranking stability, although exact top-1 prediction for the coldest items remains challenging.

The strict zero-history analysis further defines the boundary of the proposed method. When no user interaction history is available, K-RagRec performs better because item-side KG facts can be directly verbalized into the prompt. In contrast, GARDRec is designed to condition graph evidence on at least limited user behavior through temporal centroids, history-conditioned neighborhood retrieval, and interaction-aware matching. Therefore, GARDRec is best viewed as a graph-grounded LLM reranker for sparse-but-not-empty recommendation histories, while pure item-only cold-start recommendation may require an adaptive KG prompting fallback. Overall, these analyses support a bounded conclusion: graph evidence is most effective when it not only enriches the LLM input, but also participates in user-context construction, candidate comparison, and final score calibration.

\section{Conclusion and Future Work}

In this paper, we proposed \emph{GARDRec}, a graph-grounded adaptive reasoning and decision-aware framework for LLM-based next-item recommendation. Unlike existing graph-augmented LLM recommenders that mainly use knowledge graphs as prompt-level evidence, GARDRec treats graph knowledge as a decision-level grounding layer and integrates semantic-structural item representations, personalized graph contexts, inter-candidate comparison, explicit matching features, and generative calibration into a unified reranking framework. Experiments on three public benchmarks with multiple LLM backbones show that GARDRec generally outperforms representative graph-augmented LLM baselines. Ablation and diagnostic analyses further confirm the importance of graph representation, alignment between graph representations and the LLM hidden space, retrieved neighborhood context, and decision-level scoring. These results support our main claim that graph evidence should not only enrich the LLM input, but also participate in candidate comparison and final ranking decisions.

In future work, we plan to extend GARDRec from reranking to a complete retrieval and reranking pipeline. We will also explore adaptive fallback strategies for pure cold-start users and investigate more flexible graph context construction methods for domains with different metadata and knowledge-graph structures.

\printcredits

\bibliographystyle{cas-model2-names}

\bibliography{cas-refs,refs}


\vskip3pc

\end{document}